\documentclass[aps,prd,reprint,showpacs,
superscriptaddress,
nobibnotes,nofootinbib]{revtex4-1}

\usepackage{epsfig}
\usepackage{amssymb} 
\usepackage{amsmath}
\begin{document}

\title{Limit on Lorentz and CPT Violation of the Bound Neutron Using a Free 
Precession $^{3}$He/$^{129}$Xe Co-magnetometer}

\author{C. Gemmel}
\author{W. Heil}
\email[Corresponding author: ]{wheil@uni-mainz.de}
\author{S. Karpuk} 
\author{K. Lenz}
\author{Yu. Sobolev} 
\author{K. Tullney}
\affiliation{Institut f\"{u}r Physik, Johannes Gutenberg-Universit\"{a}t, 55099 Mainz, Germany}

\author{M. Burghoff}
\author{W. Kilian} 
\author{S. Knappe-Gr\"{u}neberg} 
\author{W. M\"{u}ller} 
\author{A. Schnabel} 
\author{F. Seifert} 
\author{L. Trahms}
\affiliation{Physikalisch-Technische-Bundesanstalt (PTB) Berlin, 10587 Berlin,
Germany}

\author{U. Schmidt}
\affiliation{Physikalisches Institut, Universit\"{a}t Heidelberg, 69120 Heidelberg, Germany}

\date{\today}

\begin{abstract}
We report on the search for Lorentz-violating sidereal variations of the frequency 
difference of co-located spin species while the Earth and hence the laboratory 
reference frame rotates with respect to a relic background field. The 
comagnetometer used is based on the detection of freely precessing nuclear spins from 
polarized $^{3}$He and $^{129}$Xe gas samples using SQUIDs as low-noise magnetic 
flux detectors. As result we can determine the limit for the equatorial component of the 
background field interacting with the spin of the bound neutron to be $\tilde{b}^{\text{n}}_{\bot} < 3.7 \cdot 10^{-32}$ GeV (95\%\,\,C.L.).
\end{abstract}


\maketitle

A great number of laboratory experiments have been designed to detect diminutive violations of Lorentz invariance. Among others, the Hughes-Drever-like experiments \cite{Hughes,Drever} have been performed to search for anomalous spin coupling to an anisotropy in space using electron and nuclear spins with steadily increasing sensitivity \cite{Prestage,Brown,Lamoreaux,Lamoreaux2,Chupp,Berglund,Bear,Bear2,Phillips,Cane,Heckel,Altarev}. Lorentz-violating theories should generally predict the existence of privileged reference systems. In contrast with the situation at the end of the 19th century, we have a rather unique choice nowadays for such a "preferred inertial frame," i.e., the frame where the Cosmic Microwave Background (CMB) looks isotropic. Trying to measure an anomaleous coupling of spins to a relic background field which permeates the Universe and points in a preferred direction in spacetime as a sort of new aether wind is a modern analogue of the original Michelson-Morley experiment.

The theoretical framework presented by Kosteleck\'{y} and colleagues parametrizes the general treatment of \textit{CPT}- and Lorentz violating effects in a standard model extension (SME) \cite{Colladay}. The SME was conceived to facilitate experimental investigations of Lorentz and \textit{CPT} symmetry, given the theoretical motivation for violation of these symmetries. Although Lorentz-breaking interactions are motivated by models such as string theory \cite{Kostelecky2,Ellis}, loop quantum gravity \cite{Sudarsky,Nicolai,Gambini,Crichigno}, etc. (i.e., fundamental theories combining the standard model with gravity), the low-energy effective action appearing in the SME is independent of the underlying theory. Each term in the effective theory involves the expectation of a tensor field in the underlying theory. These terms are small due to Planck-scale suppression and, in principle, are measurable in experiments. Predictions for parameters in the SME for a loop quantum gravity system with a preferred frame were discussed, e.g., in Ref. \cite{Vucetich}.

The SME contains a number of possible terms that couple to the spins of standard model particles like the electron, proton, and nucleon (mostly the bound neutron) \cite{Kostelecky}. These terms have set the most stringent limits on \textit{CPT} and Lorentz violations. To determine the leading-order effects of a Lorentz violating potential \textit{V}, it suffices to use a non-relativistic description for the particles involved given by \cite{Kostelecky}
\begin{equation} \label{GrindEQ__1_} 
V=-\tilde{b}_{J}^{w} \cdot \sigma _{J}^{w} \,\,\,\,\,\,\,\, (\text{with  } J~=~X,Y,Z~;~~~ w~=~e, p, n)~. 
\end{equation} 
The most sensitive tests were performed using a $^{3}$He-$^{129}$Xe Zeeman maser to place an upper limit on the neutron coupling to the anomalous field of $\tilde{b}_{\bot }^{n} =\sqrt{\left(\tilde{b}_{X}^{n} \right)^{2} +\left(\tilde{b}_{Y}^{n} \right)^{2} } <\, \, 10^{-31} \, \text{GeV}$ \cite{Bear,Bear2}
\noindent and, recently, by use of a K-$^{3}$He co-magnetometer thereby  improving the previous limit by a factor of 30 \cite{Brown}. An essential assumption in these so-called clock comparison experiments is that the anomalous field $\tilde{b}_{J}^{w} $ does not couple to magnetic moments but directly to the sample spins $\sigma_{J}^{w}$. This direct coupling allows comagnetometry that uses two different spin species to distinguish between a normal magnetic field and an anomalous field coupling.

The comagnetometer used for the presented measurements is based on the detection of freely spin precessing nuclear spins from polarized $^{3}$He and $^{129}$Xe samples gas with SQUIDs as low-noise magnetic flux detectors. Like in \cite{Bear,Bear2}, we search for  sidereal variations of the frequency of colocated spin species while the Earth and hence the laboratory reference frame rotates with respect to a relic background field. The observable to trace possible tiny sidereal frequency modulations is the combination of measured Larmor frequencies given by
\begin{equation} \label{GrindEQ__2_} 
\Delta \omega =\omega _{L,He} -\frac{\gamma _{He} }{\gamma _{Xe} } \cdot \omega _{L,Xe}.         
\end{equation} 

By that measure the Zeeman term is eliminated and thus any dependence on fluctuations and drifts of the magnetic field. For the $^{3}$He/$^{129}$Xe gyromagnetic ratios we took the literature values \cite{International,Pfeffer} given by $\gamma _{He} /\gamma _{Xe} =2.75408159(20)$.

The essential difference, in particular, from \cite{Bear,Bear2}, is that by monitoring the free spin precession, an ultrahigh sensitivity can be achieved with a clock which is almost completely decoupled from the environment. The design and operation of the two-species $^{3}$He/$^{129}$Xe comagnetometer has been shown recently \cite{Gemmel}. Briefly, in our measurements, we used a low-$T_{c}$ DC-SQUID magnetometer system inside the strongly magnetically shielded room BMSR-2 at PTB \cite{Bork}. A homogeneous guiding magnetic field $B_{0}$ of about 400\:nT was provided by one of the two square coil pairs which were arranged perpendicular to each other in order to manipulate the sample spins, e.g., $\pi$/2 spin flip by nonadiabatic switching. The maximum field gradients were about 33\:pT/cm. The $^{3}$He/$^{129}$Xe nuclear spins were polarized outside the shielding by means of optical pumping. Low-relaxation spherical glass vessels (R=3\:cm) were filled with the polarized $^{3}$He/$^{129}$Xe gases and placed directly below the Dewar as close as possible to SQUID sensors, which detect a sinusoidal change in magnetic flux due to the spin precession of the gas atoms in the glass cell. In order to obtain a high common mode rejection ratio, gradiometric sensor arrangements are commonly used. For our analysis it was sufficient to use a first-order gradiometer in order to suppress environmental disturbance fields. 

Nitrogen was added as buffer gas to suppress the van der Waals spin relaxation of $^{129}$Xe \cite{Chann}. In the regime of motional narrowing, i.e., at gas pressures of order mbar and at low magnetic fields \cite{Cates,Kilian}, transverse spin-relaxation times $T_{2}^{*}$ of up to 60\:h have been measured for $^{3}$He. The actual limitation in the $T_{2,\text{Xe}}^{*} $ of xenon is given by the relatively short wall relaxation time of 8\:h $<$ $T_{1,\text{wall}}^{\text{Xe}} $ $<$ 16\:h. Therefore, the total observation time $T$ of free spin-precession of our $^{3}$He/$^{129}$Xe comagnetometer is set by this characteristic time constant. According to the Cramer-Rao Lower Bound (CRLB) \cite{Kay}, the accuracy by which the frequency of a damped sinusoidal signal can be determined is given by
\begin{equation} \label{GrindEQ__3_} 
\sigma _{f} \ge \frac{\sqrt{12} }{\left(2\pi \right)\cdot SNR\cdot \sqrt{f_{BW} } \cdot T^{3/2} } \times \sqrt{C\left(T,T_{2}^{*} \right)} \, \, \, . 
\end{equation} 

\noindent $SNR$ denotes the signal-to-noise ratio, $f_{BW}$ the bandwidth, and $C(T,T_{2}^{*})$ describes the 
effect of exponential damping of the signal amplitude with $T_{2}^{*} $. For 
observation times $T \leq T_{2}^{*}$, $C(T,T_{2}^{*}$) is of order one. Deviations from the CRLB power law, due to noise sources inherent in the comagnetometer itself, did not show up in Allan standard deviation plots used to identify the power-law model for the phase noise spectrum of our runs with $T~\approx~14\:\text{h}$, typically \cite{Gemmel}. 
\begin{figure}
 \includegraphics[width=3.4in]{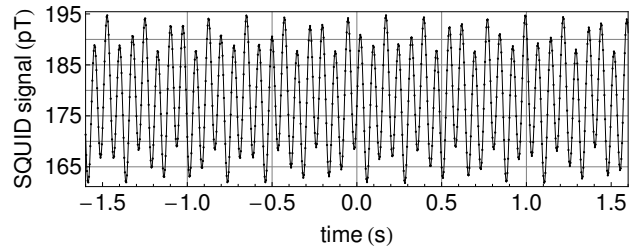}
 \caption{\label{Dataset}Typical subdata set of 3.2\:s length showing the recorded SQUID gradiometer signal of the precessing $^{3}$He/$^{129}$Xe sample spins (sampling rate: $r_{s}$=250\:Hz). The uncertainty at each data point is $\pm$ 34\:fT (\textit{k}=1) and therefore less than the symbol size. The signal amplitudes at the beginning of each run were typically $S_{\text{He}}\approx$13\:pT and $S_{\text{Xe}}\approx$4\:pT.}
\end{figure}

The  recorded signal is a superposition of the $^{3}$He and $^{129}$Xe 
precession signals at Larmor frequencies $\omega_{\text{He}} =\gamma_{\text{He}} \cdot 
B_{0} \approx 2\pi \cdot 13.4$\:Hz and $\omega_{\text{Xe}} =\gamma_{\text{Xe}} \cdot B_{0} 
\approx 2\pi \cdot 4.9$\:Hz as shown in FIG.~\ref{Dataset}. Analogue to similar problems of data analysis \cite{Hoyle2} phases of subdata sets were analyzed: The data of each run ($j$ = 1, \dots ,7) were divided into sequential time intervals ($i$) of $\tau$~=~3.2\:s ($i$ = 1, \dots ,$N_{j}$). 
The number of obtained subdata sets laid between  13350 $<$ $N_{j}$ $<$ 18000 corresponding to 
observation times $T_{j}$ of coherent spin precessions in the range of 12\:h $<$ $T_{j}$ $<$ 16\:h.
For each subdata set a $\chi^{2}$ minimization was performed, using the fit function 
\begin{align} \label{GrindEQ__4_} 
\nonumber A^{i} (t) = &A_{\text{He}}^{i} \cdot \sin \left(\omega _{\text{He}}^{i} t\right)+
B_{\text{He}}^{i} \cdot \cos \left(\omega _{\text{He}}^{i} t\right)+ \\
\nonumber & A_{\text{Xe}}^{i} \cdot \sin \left(\omega _{\text{Xe}}^{i} t\right)+
 B_{\text{Xe}}^{i} \cdot \cos \left(\omega _{\text{Xe}}^{i} t\right)+ \\
 & (c_{0}^{i} +c_{\text{lin}}^{i} \cdot t) 
\end{align} 

\noindent with a total of 8 fit parameters. Within the relatively short time intervals, 
the term ($c_{0}^{i} + c_{\text{lin}}^{i} \cdot t$) represents the adequate parameterization 
of the SQUID gradiometer offset showing a small linear drift due to the elevated 1/$f$ noise at low frequencies 
($<$ 1\:Hz) \cite{Gemmel}. 
On the other hand, the chosen time intervals are long enough to have a sufficient number of data points (800)
for the $\chi^{2}$ minimization. The sum of sine and cosine terms are chosen to have only linear fitting 
parameters for the subdata set phases which are given by
\begin{equation} \label{GrindEQ__5_} 
\varphi^{i} =\arctan \, \, (B^{i} /A^{i})~.
\end{equation} 
The normalized $\chi^{2}$ ($\chi^{2}$/d.o.f) of most subdata sets ($i$) is close 
to 1 which is consistent with the assigned uncertainty to each data point of 
$\pm$\:34\:fT (\textit{k}=1); see FIG.~\ref{Dataset}. The latter value is the typical noise signal $N_{s}$ 
derived from the mean system noise $\bar{\rho }_{s} \approx 3\:\text{fT}/ \sqrt{\text{Hz}}$ in the recorded effective bandwidth of 100\:Hz. 
Jumps in the SQUID signal in the order of 1\:pT caused by external disturbances 
gave $\chi^{2}$/d.o.f\:$\gg$\:1 for the respective subdata sets. In the analysis we therefore disregarded all subdata sets with $\chi^{2}$/d.o.f\:$\geq$\:2 ($<$\:0.5\% 
in total). 
\noindent For each subdata set of chosen time interval 
-1.6\:s\:$\leq\:(t-t_{i-1,j})\:\leq$\:+1.6\:s (see FIG.~\ref{Dataset}), we finally obtain numbers 
for the respective fit parameters $\omega_{\text{He}}^{i},\: \omega_{\text{Xe}}^{i},\: \varphi_{\text{He}}^{i},\: 
\varphi_{\text{Xe}}^{i}$ including error bars. 

\begin{figure}
\begin{center}
\begin{minipage}{0.49\textwidth}
 \includegraphics[width=3.65in]{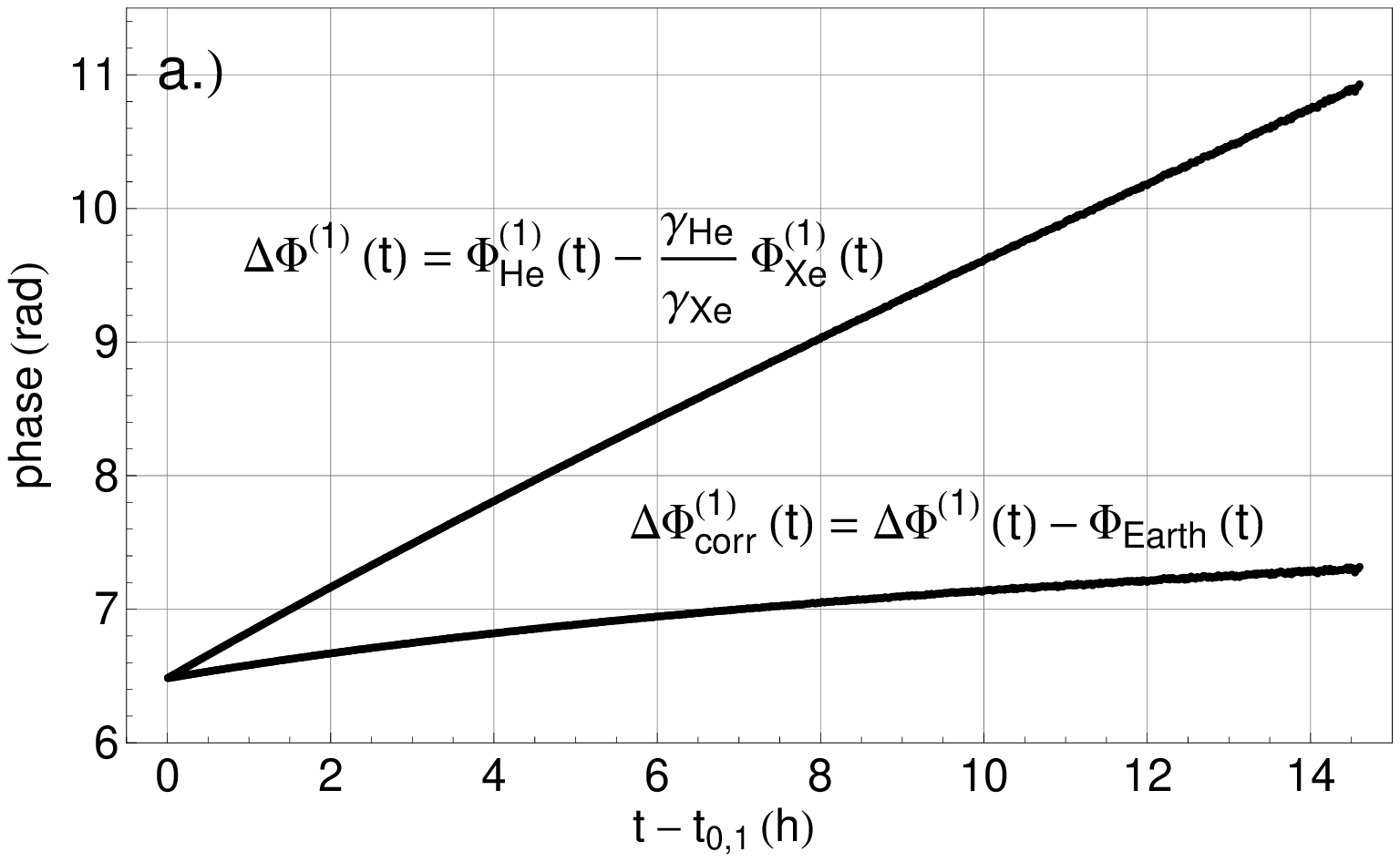}
\end{minipage}
\begin{minipage}{0.49\textwidth}
 \includegraphics[width=3.65in]{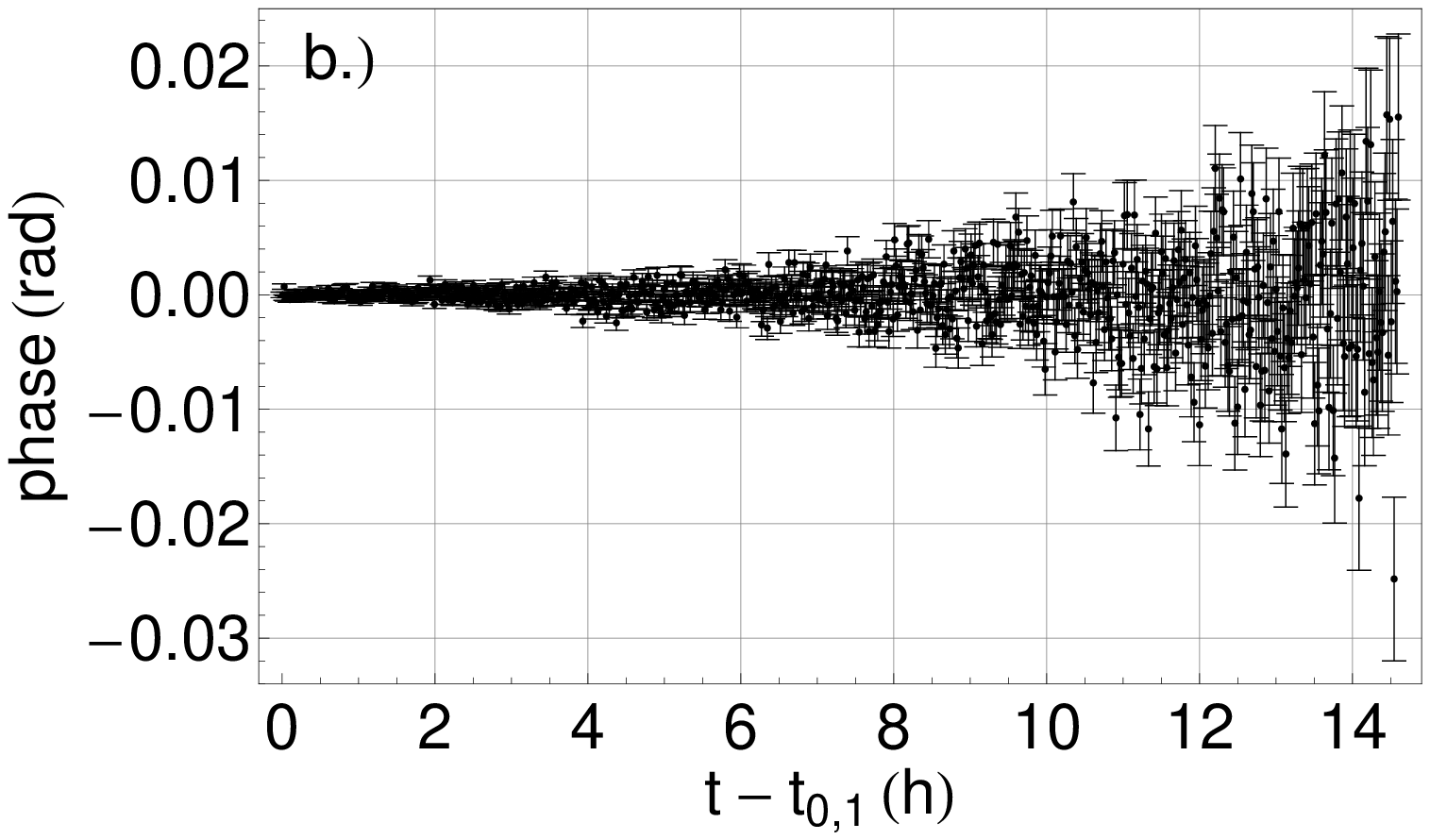}
 \caption{\label{phasesresiduals}(a) Measured phase differences 
$\Delta \Phi ^{(1)}(t)$ for run 1 and the corresponding corrected phases $\Delta \Phi_{\text{corr}}^{(1)}(t)$ after subtraction of the effect of the Earth's rotation. 
(b) Phase residuals after subtraction of phase drifts given by the fit model of Eq.~(\ref{GrindEQ__7_}) (one data point 
comprises 20 subdata sets, i.e., $\Delta t=64$~s).} 
\end{minipage}
\end{center}
\end{figure}

In a further step, we can deduce values for the average frequency 
$\bar{\omega }^{j} =\frac{1}{N_{j}} 
\sum _{i=1}^{N_{j}}\omega^{i}$ for each run. The accumulated phase 
(omitting the index $j$) is then determined to be\footnote{As the maximal 
frequency deviation $\Delta \omega$ from the mean $\bar{\omega }^{j}$  
was smaller than $5 \cdot 10^{-6}$\:rad/s in the course of one run \cite{Gemmel}, 
we had at all times $\Delta \omega \cdot \tau \ll 2 \pi$.}
\begin{eqnarray} \label{GrindEQ__6_} 
\nonumber \Phi \left(t=m\tau \right) &=& \Phi \left(t=(m-1)\tau \right)+ \bar{\omega} \cdot \tau + \varphi^{m} \nonumber \\
                                     & & - \bmod [\Phi \left(t=(m-1)\tau \right)+ \bar{\omega} \tau;~ 2\pi ]   
\end{eqnarray} 
\noindent with $m$=1,...,N-1 and $\Phi(t=0)=\varphi^{1}$ being the phase offset of the first time interval.
\noindent Following Eq.~(\ref{GrindEQ__2_}) the extracted phase difference 
$\Delta \Phi^{(1)} \left(t\right)=\Phi_{\text{He}}^{(1)} 
\left(t\right)-\left(\gamma_{\text{He}} /\gamma_{\text{Xe}} \right)\cdot 
\Phi_{\text{Xe}}^{(1)}\left(t\right)$ is plotted for run 1. $\Delta \Phi^{(1)} \left(t\right)$ 
is expected to be constant if there is no sidereal modulation of the spin-precession frequency and/or no other drifts and noise sources. Nevertheless, in addition 
to an arbitrary phase offset an almost linear time dependence is seen in FIG.~\ref{phasesresiduals}(a). The dominant 
contribution is caused from the Earth's rotation, i.e., the rotation of the SQUID 
detector with respect to the precessing spins. For the location of the PTB Berlin, Germany ($\theta$=52.5164$^{\circ}$\:north) and the angle between north-south direction and the guiding magnetic field ($\rho$=28$^{\circ}$), the linear term in the weighted phase difference due to Earth's rotation is given by \cite{Gemmel}
\begin{eqnarray} \label{phase_earth}
\Phi _{\text{Earth}} & = & -\Omega_{\text{SD}} \cdot (1-\gamma_{\text{He}}/\gamma_{\text{Xe}})\cdot \cos \rho \cdot \cos \theta \cdot t \nonumber \\
              & = & 6.87263\times 10^{-5} \mbox{rad/s} \, \cdot t~. 
\end{eqnarray}
\noindent$\Omega_{\text{SD}}$ is the angular frequency of the sidereal day with $\Omega_{\text{SD}}=2\pi/T_{\text{SD}}=2\pi/(23^{\text{h}}:\; 56^{\text{min}}:\; 4.091^{\text{s}})$. Subtracting this term from $\Delta \Phi^{(1)}(t)$, we get the corrected phase $\Delta \Phi_{\text{corr}}^{(1)}(t)$ which is plotted in FIG.~\ref{phasesresiduals}(a), too. Let us assume that there is no sidereal variation of the 
$^{3}$He/$^{129}$Xe frequencies induced by Lorentz-violating couplings, and then
$\Delta \Phi_{\text{corr}}^{(j)}(t)$ can be described best by 
\begin{eqnarray} 
\label{GrindEQ__7_} 
\Phi_{\text{fit}}^{(j)}\left(t\right)&=&\Phi_{0}^{(j)}+\Delta\omega_{\text{lin}}^{(j)}\left(\,t-t_{0,j}\right) \nonumber \\
&&+E_{\text{He}}^{(j)}\cdot\exp\,\left(\frac{-\left(\,t-t_{0,j}\right)}{T_{2,\text{He}}^{*\, (j)}}\right) \nonumber \\
&&+E_{\text{Xe}}^{(j)} \cdot \exp \, \left(\frac{-\left(\, t-t_{0,j} \right)}{T_{2,\text{Xe}}^{*\, (j)} } \right)  
\end{eqnarray} 
\noindent with  
\begin{equation}
\Phi _{\text{fit}}^{(j)} (t)=\left\{\begin{array}{l} {\Phi _{\text{fit}}^{(j)}(t)\, \, \, \, \mbox{for}\, \, \, t_{0,j} \le t\le (t_{0,j} +N_{j} \cdot \tau)} \\ {0\, \, \, \, \, \, \, \, \, \, \, \, \, \, \, \, \, \, \,
\mbox{elsewhere.}\, \, \, \, \, \, \, \, \, \, \, \, \, \, \, } \end{array}\right. 
\nonumber
\end{equation}
\noindent $t_{0,j}$ is the absolute starting time of each run. Our interpretations of the terms are as follows: $\Phi_{0}^{(j)}$ is a 
general phase offset and $\Delta \omega_{\text{lin}}^{(j)}(t-t_{0,j})$ is an additional 
linear phase shift mainly arising from deviations of the gyromagnetic ratios 
of $^{3}$He and $^{129}$Xe from their literature values due to chemical shifts and uncertainties in the subtraction of $\Phi_{\text{Earth}}$ \cite{Gemmel}. The 
two exponential terms with amplitudes $E_{\text{He}}^{(j)}$ and $E_{\text{Xe}}^{(j)}$ reflect 
the respective phase shift due to demagnetization fields in a nonideal 
spherical cell seen by the spin ensembles (self-shift). These phase shifts are 
directly correlated to the decay times $T_{2,\text{He}}^{(j)}$ and $T_{2,\text{Xe}}^{(j)}$ of the respective signal amplitude of the precessing 
helium and xenon spins \cite{Gemmel}. As the $T_{2}^{*(j)}$ times can be determinded independently for both spin species from the experiment, four 
fit parameters are left for each run, such that the fit model is basically a linear 
function in parameters.

Fitting the corrected phase difference $\Delta \Phi_{\text{corr}}^{(j)}(t)$ to Eq.~(\ref{GrindEQ__7_}) 
and subtracting the fit function from $\Delta \Phi_{\text{corr}}^{(j)}(t)$ results in the phase residual as shown 
for run 1 in FIG.~\ref{phasesresiduals}(b). Because of the exponential decay of the signal 
amplitudes, mainly that of xenon with the shorter $T_{2,\text{Xe}}^{*}$ of only 4-5\:h, the 
$SNR$ decreases resulting in an increase of the residual phase noise, i.e., 
$\sigma_{\Phi,\text{res}} \propto \exp\left(t/T_{2,\text{Xe}}^{*(j)}\right)$ \cite{Gemmel}.

In the last step, a piecewise fit function was defined, which is a combined fit to all seven runs, now including the parameterization of the sidereal phase modulation
\begin{figure}
 \includegraphics[width=3.45in]{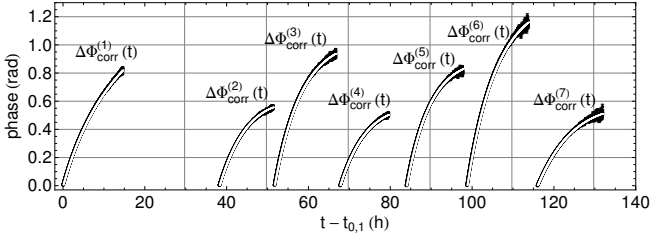}
 \caption{\label{allphases}Corrected phase differences $\Delta \Phi_{\text{corr}}^{(j)}(t)$ 
with combined fit function $\Phi_{\text{fit}}^{\text{SD}}(t)$ (white solid line) for the seven runs (one data point 
comprises 20 sub-data sets, i.e., $\Delta t=64$~s). In order to present these results in a common plot, the general phase offset $\Phi_{0}^{(j)}$ was subtracted from $\Delta \Phi_{\text{corr}}^{(j)}(t)$ for each run.} 
\end{figure}
\begin{eqnarray} \label{GrindEQ__8_} 
\nonumber \Phi_{\text{fit}}^{\text{SD}}(t)&=&\sum_{j=1}^{7} \Phi_{\text{fit}}^{(j)}(t)\\
\nonumber                     & & +\{\,a_{\text{s}}\cdot \sin \left(\Omega_{\text{SD}} \left(t-t_{0,1} \right)+\varphi_{\text{SD}} \right) \\
                              & & -\,\,a_{\text{c}} \cdot \cos \, \left(\Omega_{\text{SD}} \left(t-t_{0,1} \right)+\varphi_{\text{SD}}\right)\,\} 
\end{eqnarray} 
$\varphi_{\text{SD}}=2 \pi \cdot t_{\text{SD}}$ represents the phase offset of sidereal modulation at the local sidereal time (vernal equinox J2000.0) on 2009 March 21 at 20:52 UT (universal time) which is the starting time $t_{0,1}$ of the first run. Neglecting multiples of 24\;h the local sidereal time is 9.7\;h, which in units of sidereal day gives $t_{\text{SD}} =\, \, 0.4053$.

\begin{table*}
\begin{center}
\begin{minipage}{4.7in}
\caption{Results for the sidereal phase amplitudes $a_{\text{c}}$ and $a_{\text{s}}$ together with their correlated and uncorrelated 
1$\sigma$ errors (2nd row) determined by a $\chi^{2}$ minimization using the fit model of Eq.~(\ref{GrindEQ__8_}). In order to demonstrate the strong dependence of 
the correlated error on the angular frequency of the sidereal day $\Omega_{\text{SD}}$, corresponding fit results are shown 
for multiples of $\Omega_{\text{SD}}$: $\Omega_{\text{SD}}^{'} = g \cdot \Omega_{\text{SD}}$.\vspace{0.05in}}
\end{minipage}
\begin{tabular}{lcccccc}
\hline
\hline
 & $a_{\text{c}} \text{(mrad)} $ & $\sigma _{a_{\text{c}}}^{\text{corr}} \text{(mrad)} $ & $\sigma_{a_{\text{c}}}^{\text{uncorr}} \text{(mrad)} $ & $a_{\text{s}} \text{(mrad)} $ & $\sigma_{a_\text{s}}^{\text{corr}} \text{(mrad)} $ & $\sigma_{a_{\text{s}}}^{\text{uncorr}} \text{(mrad)} $ \\
\hline
       0.5$\cdot \Omega_{\text{SD}}$    &    3.353  &       6.572 &    0.018 &      0.488  &   7.991 &    0.016 \\
       $\Omega_{\text{SD}}$             &    -0.882 &       0.814 &    0.015 &      -2.067 &   1.057 &    0.019 \\
       2$\cdot \Omega_{\text{SD}}$      &    -0.048 &       0.120 &    0.016 &      -0.149 &   0.112 &    0.017 \\
       3$\cdot \Omega_{\text{SD}}$      &    -0.184 &       0.052 &    0.019 &      -0.011 &   0.043 &    0.016 \\
       4$\cdot \Omega_{\text{SD}}$      &    -0.001 &       0.034 &    0.018 &      0.057  &   0.030 &    0.016 \\
\hline
\hline
\end{tabular}  
\label{tabelle}
\end{center}
\end{table*}

Figure~\ref{allphases} shows the corrected phase differences 
$\Phi_{\text{corr}}^{(j)}(t)$ together with the fit function 
$\Phi_{\text{fit}}^{\text{SD}}(t)$ (white solid line) for all seven runs. The $\chi^{2}$/d.o.f of the fit gave 
1.868, which shows that the phase model of Eq.~(\ref{GrindEQ__7_}) may be somewhat 
incomplete or, what is more likely, the phase errors are underestimated in the analysis of the subdata sets. In order to take an (unknown) uncertainty into account, the errors on the phases were scaled to obtain a $\chi^{2}$/d.o.f of one, as recommended, e.g., by Refs.\:\cite{Particle,Williams}. In Table \ref{tabelle} (2nd row) the fit results for the amplitudes $a_{\text{c}}$ and $a_{\text{s}}$ of the sidereal phase modulation are shown together with their correlated and uncorrelated 
1$\sigma$ errors. 

It is noticeable that the uncorrelated error 
which represents the integrated measurement sensitivity of our $^{3}$He/$^{129}$Xe 
comagnetometer is about a factor of 50 less than the correlated one. The 
big correlated error on $a_{\text{s}}$ and $a_{\text{c}}$ is caused by a piecewise similar 
time structure of $\Phi_{\text{fit}}^{(j)}(t)$ and the sidereal phase modulation in the fit function of Eq.~(\ref{GrindEQ__8_}). 
On a closer look, this can be traced back to the relatively short $T_{2,\text{Xe}}^{*(j)}$ times (compared to $T_{\text{SD}}$) that enter in the argument of the exponential terms of Eq.~(\ref{GrindEQ__7_}). Therefore, the present sensitivity limit of our $^{3}$He/$^{129}$Xe comagnetometer is set by the correlated error. In order to substantiate that more clearly, we changed the fit model of Eq.~(\ref{GrindEQ__8_}) by taking multiples of $\Omega_{\text{SD}}$ ($\Omega_{\text{SD}}^{'} =g\cdot \Omega_{\text{SD}}$), 
i.e., replacing $T_{\text{SD}}$ by $T^{'}_{\text{SD}} =T_{\text{SD}}/g$. The results show that the correlated error approaches the uncorrelated one already for $g \geq$ 3 (see Table\:\ref{tabelle}). 
The uncorrelated error, however, is only marginally 
affected by this procedure, as expected. From Table~\ref{tabelle} (2nd row) we now extract the rms value of the sidereal phase amplitude $\Phi_{\text{SD}} =\sqrt{a_{\text{s}}^{2} +a_{\text{c}}^{2}}$, yielding 
(2.25\:$\pm$\:2.29)\:mrad (95\%\:C.L.). This result is consistent with the abscence of Lorentz- and 
\textit{CPT}-violating effects, giving reasonable assumptions about the probability 
distribution for $\Phi_{{\text{SD}}}$ \cite{explanation}. 

\noindent In terms of the SME \cite{Kostelecky} we can express the sidereal phase amplitudes according to
\begin{equation}
a_{\text{s}} =\frac{2\pi }{\Omega _{\text{SD}} } \cdot \delta \nu _{\text{X}} \:\:\: \text{and} \:\:\: a_{\text{c}} =\, \, \frac{2\pi }{\Omega _{\text{SD}} } \cdot \delta \nu _{\text{Y}} \\ \vspace{0.2cm}
\label{GrindEQ__9_}
\end{equation}
\noindent with
\begin{equation}
2\pi \left|\delta \nu _{\text{X,Y}} \right|\cdot \hbar =\left|\, 2\cdot \left(1- \gamma_{\text{He}}/\gamma_{\text{Xe}} \right)\cdot \sin \chi \cdot \tilde{b}_{\text{X,Y}}^{\text{n}} \right|~.\\ \vspace{0.2cm}
\label{GrindEQ__20_}
\end{equation}

\noindent $\chi$ is the angle between the Earth's rotation axis and the 
quantization axis of the spins with $\chi=\arccos \, \, (\cos\theta\cdot\cos\rho)=57^{\circ}$. Within the Schmidt model \cite{Schmidt}, 
the valence neutron of $^{3}$He and $^{129}$Xe determines the spin and the magnetic 
moment of the nucleus. Thus, our $^{3}$He/$^{129}$Xe comagnetometer is sensitive to the 
bound neutron parameters $\tilde{b}_{\text{X,Y}}^{\text{n}}$. From that, we can deduce numbers for $\tilde{b}_{\text{X,Y}}^{\text{n}}$:
\begin{eqnarray}
\tilde{b}_{\text{X}}^{\text{n}}     &=& (3.36\pm 1.72) \cdot 10^{-32} \; \text{GeV}\;\; (1\sigma),  \\
\tilde{b}_{\text{Y}}^{\text{n}}     &=& (1.43\pm 1.33) \cdot 10^{-32} \; \text{GeV}\;\; (1\sigma), 
\label{GrindEQ__10_}
\end{eqnarray}
\noindent which can be interpreted as $\left|\tilde{b}_{\bot }^{\text{n}}\right| < 3.7 \cdot 10^{-32}$\:GeV at 95$\%$ confidence level for the upper limit of the equatorial component of the background tensor field interacting with the spin of the bound neutron. For the calculation of the upper limit on $\tilde{b}_{\bot }^{\text{n}}$ Eq.~(\ref{GrindEQ__9_}) and Eq.~(\ref{GrindEQ__20_}) were used putting in the 95\%\:C.L. for the rms value of the sidereal phase amplitude $\Phi_{\text{SD}}$.\\

Further improvements for Lorentz and \textit{CPT} tests using the free spin-precession $^{3}$He/$^{129}$Xe comagnetometer can be achieved via two mayor steps: First, the relatively short wall relaxation time of $^{129}$Xe limiting the total observation time $T$ of free spin precession has to be increased considerably (T$_{1,\text{wall}}\:\approx\:$T$_{\text{SD}}$) such that we approach the measurement sensitivity given by the uncorrelated error. Since the latter one follows the $\propto$T$^{-3/2}$ power law according to CRLB of Eq.~(\ref{GrindEQ__3_}), the longer observation time $T$ will lead to an additional increase in sensitivity. Second, the number of measurement runs has to be increased to a period of 100 days. Besides gain in statistics, the long time span provides an important separation between sidereal and possible diurnal variations. \\

This work was supported by the Deutsche Forschungsgemeinschaft (DFG) under Contract No. BA 3605/1-1.


\begin{thebibliography}{99}
\bibitem{explanation} The probability distribution $P$ for $\Phi_{{SD}}$ 
is formed by multiplying the distributions for a$_{s}$ and a$_{c}$, converting to polar 
coordinates, and integrating over the polar angle. If a$_{s}$ 
and a$_{c}$ have zero mean value and the same error $\sigma$, the probability distribution takes the form $P(\delta \Phi _{SD} )=\frac{\Phi _{SD} }{\sigma ^{2} } \cdot \exp \, \left(\frac{-\Phi _{SD}^{2} }{2\sigma ^{2} } \right)$. 
As an approximation we take the average of both errors 
$\sigma =\left(\delta a_{s} +\delta a_{c} \right)/2$ and determine the 95\% 
error by setting $\int _{0}^{\delta \Phi _{SD} }P\left(\Phi _{SD} \right)\, 
d\Phi _{SD} =0.95 $ giving $\delta \Phi_{SD}$=2.29\:\text{mrad}.
\end{thebibliography}
\end{document}